\documentclass{article}
\usepackage{graphicx}

\begin{document}

\title{Distributed Denial of Service is a Scalability Problem}
\author{Yoo Chung}
\maketitle

\begin{abstract}
  Distributed denial of service attacks are often considered a security problem.  While this may be the way to view the problem with the Internet of today, new network architectures attempting to address the issue should view it as a scalability problem.  In addition, they need to approach the problem based on a rigorous foundation.
\end{abstract}

\section{Overwhelming systems}
\label{sec:dos}

Imagine a scene from a movie.  A hacker is living happily with his computers that fill up every corner of his house.  Then one day, a soda manufacturer decides to change the formula for his favorite carbonated drink, and not in a way that makes the hacker happy.  With his happiness being threatened, he decides to crack into the manufacturer's online systems, typing on his keyboard like mad, until he manages to shut down the manufacturer and lets his ire be known.

In the real world, computer security incidents do not quite happen like this.  Typically, a programmer or a team of programmers in the employ of a criminal enterprise, or even a legitimate government, will create malicious software that can be spread in the form of trojans, viruses, or worms.  This software is let loose into the wild, infecting thousands or even millions of computers, most of them exploited due to vulnerabilities in the software or poor security practices by their users.\footnote{And often both, with so many networked computers failing to apply fixes that get rid of old vulnerabilities.}  This huge collection of infected computers, called a botnet~\cite{grizzard:hotbots2007}, lies dormant until someone finds a use for it, at which point some person, who does not even necessarily possess much computer skill, leisurely instructs the botnet to fulfill whatever nefarious purpose the person wants fulfilled.\footnote{The author, not knowing anyone who uses botnets, does not really know if it is done ``leisurely'', but he is pretty sure that it does not involve mad typing skills.}

One of these nefarious purposes is to execute a distributed denial of service attack~\cite{peng:csur2007} on a specific online system so that nobody can use it.  A distributed denial of service attack involves a very large number of computers all over the network sending data to the single target system that is to be attacked.  If the target system is unable to keep up with the deluge of incoming data, then it will be unable to handle legitimate data that it is being sent, which basically is the same as shutting down the target system as far as legitimate users are concerned.  As a concrete example, if a web server is being sent an overwhelming number of web requests by a botnet, then legitimate users cannot access any content from the web server.

\section{It's scalability, not security}
\label{sec:problem-scalability}

Because distributed denial of service attacks can kill the availability of networked systems, they are considered a security problem, and the thought processes involved when trying to solve the problem are often in terms of defense, and sometimes even offense~\cite{walfish:tocs2010}, where systems attempt to detect an attack and respond correspondingly.

With the current architectures for the Internet, the World Wide Web, and many other systems built upon the Internet, there has been little choice but to either use this approach, or to make an enormous investment so that enough computing resources are available such that attackers cannot overwhelm a target system.  Only a few organizations can afford and are willing to make such huge investments for the latter, while the rest of us try to cope by leaving ourselves vulnerable to distributed denial service attacks or investing in ``security solutions'' for detecting and handling them.

Either way, it has been a case of defenders and attackers leapfrogging each other. Defenses are erected against distributed denial of service attacks, attacks are devised to overcome these defenses, yet new defenses to handle these new attacks arise, ad infinitum.  And there is no indication that this can ever stop given current network architectures.

This need not be the case when creating new network architectures.  In fact, if a new network architecture claims to be resistant to distributed denial of service attacks, then it should \emph{not} be the case.  However, it is the author's belief that if a network architecture approaches the problem as a \emph{security} problem, where it attempts to detect distributed denial of service attacks and react accordingly, then the same cycle today will merely repeat with the new architecture.

Distributed denial of service attacks are not the only situations in which online systems can be overwhelmed.  Being \emph{popular} is another way that systems get overwhelmed.  While large organizations sometimes make large investments in techniques such as load balancing or content distribution networks to handle huge loads, individuals and smaller organizations usually are not able to afford such investments.  When the demand for some of their content or services rises to a much more higher level than usual, their systems cannot keep up and eventually end up inaccessible.  The phenomenon even has a name, the ``Slashdot effect'', after how small websites would often be overwhelmed when the popular Slashdot website linked to them.

While it may be the case that illegimate traffic from distributed denial of service attacks and legitimate traffic from a huge surge in popularity have detectable differences, the differences are only going to grow smaller.  The programmers of botnet software responsible for distributed denial of service attacks do not want network operators able to filter out their attacks, so they will continue to devise ways to make distributed denial of service attacks look like legitimate traffic.  And if there ever comes the day when there is no way to distinguish distributed denial of service attacks from a huge surge in popularity, then trying to detect attacks and block them becomes a lost cause.

The network operators of today have no choice but to handle distributed denial of service attacks as a security problem.  However, those designing new network architectures should consider it a \emph{scalability} problem.  If a network architecture is able to support the operation of any service without disruption no matter any surge in network traffic, then smaller operators of networking applications such as small websites will no longer have to worry about popularity becoming a curse.  More importantly, distributed denial of service is fundamentally cut off at its knees.

Long story short: In the Internet of the present, distributed denial of service is a security problem; In creating the Internet of the future, it should be a scalability problem.

\section{Be rigorous}
\label{sec:be-rigorous}

Even when a new network architecture is designed with scalability in mind from the very start, if it is not rigorously shown to be scalable, then it would be quite likely that there is still an unscalable aspect that malicious entities could exploit in a distributed denial of service attack.  This section will describe a cautionary tale.

Content-centric networking~\cite{jacobson:conext2009} is a network architecture put forward to simplify many aspects of the network compared to the Internet of today while dramatically increasing flexibility.  The aspect that we will focus on is its potential resistance to distributed denial of service attacks.  In \cite{jacobson:conext2009}, the authors claim that data-based distributed denial of service attacks are simply not possible, and while interest-based attacks would be theoretically possible, they would also be very easy to mitigate.  While the former claim is not disputed here, the latter already faces a problem with a sort of ``push-pull'' attack.

\begin{figure}
  \centering
  \includegraphics[width=8cm]{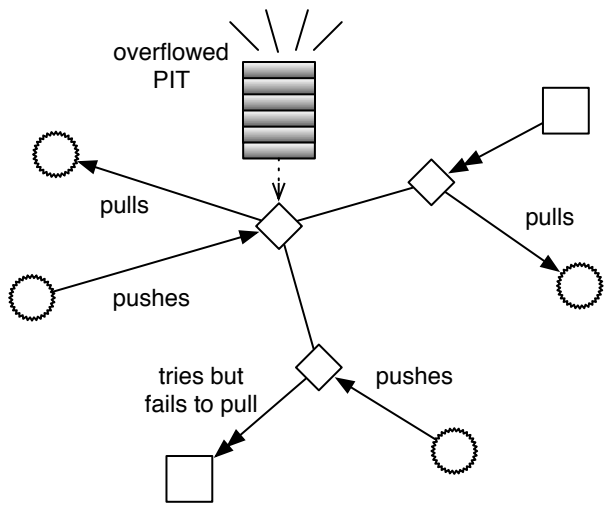}
  \caption{Push-pull attack against a content-centric network.}
  \label{fig:push-pull}
\end{figure}

With a huge botnet, a malicious entity could divide the group into ``pushers'' and ``pullers''.  The pushers willl generate a huge amount of content with distinct names.  The pullers, on the other hand, will generate a huge amount of interests which pull these content, since it should not be too difficult for the pushers and pullers to generate and request the same content names, programmatically or otherwise.  Because so many interests are being broadcast by the pullers in the botnet, it can overfill the pending interest table (PIT) in a content router, which can prevent a legitimate interest from being forwarded to where the content is, which in turn prevents the legitimate consumer from receiving the content.  It is basically an interest-based distributed denial of service attack, and the concept is shown in figure~\ref{fig:push-pull}.

Neither of the suggestions in \cite{jacobson:conext2009} that should prevent a simpler interest-based attack is sufficient to defend against this push-pull attack.  The pullers requesting content generated by the pushers prevents throttling of interests based on how many goes unfulfilled.  And a botnet by its very nature exploits vulnerable computers all over the network, so it is unlikely that filtering by name prefix would be effective.  Not only that, but given how a botnet is deployed by infecting vulnerable computers, there is no guarantee that any cryptographic protection will not be co-opted.

This attack was quickly thought up by the author after reading a technical description of content-centric networking.  If such an attack can be devised in a short time, how many more attacks can be imagined by someone who spends all their time trying to think of them?  In fact, \cite{lauinger2010} describes several potential distributed denial of service attacks against content-centric networking, including the push-pull attack described in this section.

The point is not that content-centric networking is fundamentally flawed in terms of resisting distributed denial of service attacks.  It is that if network architectures are designed in a way such that it is only resistant to attacks their designers think of, it will almost certainly be the case that other ways to execute a distributed denial of service attack will be devised.  The only way to ensure that a network architecture is resistant to such attacks is to rigorously show, ideally through mathematical proof, that no matter what happens,\footnote{Within feasible constraints, of course.  For example, destroying the Earth would be a very definitive way to deny all services, but it is hardly a feasible attack.} there is no way to overwhelm any aspect of the architecture.\footnote{In fact, the author believes that such a design could be possible for content-centric networking.}

Admittedly, even mathematical proof of a system's total scalability is not an absolute guarantee that distributed denial of service attacks can never succeed again.  Just look at the development of side-channel attacks against cryptographic systems that arose in the 1990s~\cite{kocher:crypto1996}, which sidestepped assumptions that mathematically guaranteed the security of various cryptographic algorithms.  However, such paradigm-shattering developments have got to be far rarer than incremental advances: with a rigorous foundation, a network architecture could be immune to distributed denial of service attacks for a decade or two, rather than new attacks overcoming new defenses practically every year.

\section{Concluding thoughts}
\label{sec:conclude}

It looks like the architecture underlying the Internet and the World Wide Web today has aspects that are fundamentally not scalable.  This will most likely mean an endless cycle of new distributed denial of service attacks, defenses that protect against these attacks, attacks that plow through these defenses, yet more defenses, over and over again.

For brand new networking architectures that start from a clean slate, this would not necessarily be the case.  A network architecture that is scalable in all aspects would be fundamentally immune to distributed denial of service attacks.  It has to be rigorously shown to be scalable, however.  Otherwise, it is likely that the same cycle that the Internet goes through today will just repeat in any future Internet.  If a totally scalable network architecture is proven impossible to exist, i.e. a ``full employment theorem for the distributed denial of service field'', then we must be prepared to live with this unending cycle.  The author hopes that reality is more forgiving, and that such a network architecture can indeed exist.

Finally, the author wishes to point out that a totally scalable network architecture will not render obsolete all research related to detecting and reacting to distributed denial of service attacks.  Such attacks may no longer be ``denial of service'' attacks in the future, but they may well still be ``cost-increasing'' attacks by consuming more bandwidth, processing power, or storage than would otherwise be required.  This may not be as sensational as shutting down major websites, but they would still be potential extra costs that could be used for extortion.

\bibliographystyle{plain}
\bibliography{strings,local,articles,proceedings}

\begin{thebibliography}{1}

\bibitem{grizzard:hotbots2007}
Julian~B. Grizzard, Vikram Sharma, Chris Nunnery, Brent~ByungHoon Kang, and
  David Dagon.
\newblock Peer-to-peer botnets: Overview and case study.
\newblock In {\em Proceedings of the First Workshop on Hot Topics in
  Understanding Botnets}. USENIX, April 2007.

\bibitem{jacobson:conext2009}
Van Jacobson, Diana~K. Smetters, James~D. Thornton, Michael~F. Plass,
  Nicholas~H. Briggs, and Rebecca~L. Braynard.
\newblock Networking named content.
\newblock In {\em Proceedings of the 5th International Conference on Emerging
  Networking Experiments and Technologies}, pages 1--12. ACM Press, December
  2009.

\bibitem{kocher:crypto1996}
Paul~C. Kocher.
\newblock Timing attacks on implementations of {D}iffie-{H}ellman, {RSA},
  {DSS}, and other systems.
\newblock In {\em Proceedings of the 16th Annual International Cryptology
  Conference on Advances in Cryptology}, pages 104--113. Springer-Verlag,
  August 1996.

\bibitem{lauinger2010}
Tobias Lauinger.
\newblock Security \& scalability of content-centric networking.
\newblock Master's thesis, Technische Universit{\"{a}}t Darmstadt and
  Eur{\'{e}}com, September 2010.

\bibitem{peng:csur2007}
Tao Peng, Christopher Leckie, and Kotagiri Ramamohanarao.
\newblock Survey of network-based defense mechanisms countering the {DoS} and
  {DDoS} problems.
\newblock {\em ACM Computing Surveys}, 39(1), April 2007.

\bibitem{walfish:tocs2010}
Michael Walfish, Mythili Vutukuru, Hari Balakrishnan, David Karger, and Scott
  Shenker.
\newblock {DDoS} defense by offense.
\newblock {\em ACM Transactions on Computer Systems}, 28(1), March 2010.

\end{thebibliography}

\end{document}